\documentclass[prd,twocolumn,amsmath,amssymb,floatfix, superscriptaddress]{revtex4}
\usepackage{graphicx, epsfig, bm,amsmath,natbib}

\usepackage{color}
\usepackage{hyperref}
\usepackage{ifthen}
\usepackage{xstring}
\usepackage{graphicx,epsfig}
\usepackage{subfigure}
\usepackage{amssymb}
\usepackage{afterpage}

\def\be{\begin{equation}}
\def\ee{\end{equation}}
\def\ba{\begin{eqnarray}}
\def\ea{\end{eqnarray}}
\def\nn{\nonumber}

\begin{document}

\title{CMB Constraints on Principal Components of the Inflaton Potential}

\author{Cora Dvorkin}
\affiliation{Kavli Institute for Cosmological Physics, Enrico Fermi Institute,
        University of Chicago, Chicago, IL 60637}
\affiliation{Department of Physics, University of Chicago, Chicago, IL 60637}

\author{Wayne Hu}
\affiliation{Kavli Institute for Cosmological Physics, Enrico Fermi Institute,
        University of Chicago, Chicago, IL 60637}
\affiliation{Department of Astronomy \& Astrophysics, University of Chicago, Chicago, IL 60637}

\begin{abstract}
We place functional constraints on the shape of the inflaton potential from the cosmic microwave background through a variant of the generalized slow roll approximation that allows large amplitude, rapidly changing deviations from scale-free conditions.
Employing a principal component decomposition of the source function 
$G' \approx 3 (V'/V)^2 - 2 V''/V$ and keeping only those measured to better than
$10\%$ results in 5 nearly independent Gaussian constraints that may
be used to test any single-field inflationary model where such deviations are expected.
The first component implies   $< 3\%$ variations at the 100 Mpc scale.  One component
shows a 95\% CL preference for deviations around the 300 Mpc scale at the 
$\sim 10\%$ level but the global significance is reduced considering the 5 components examined.  This deviation also requires a change in the cold dark matter density which in a flat $\Lambda$CDM model
is disfavored by current supernova and Hubble constant data and can be tested with future polarization or high multipole temperature
data.  Its impact resembles a local running of the tilt from multipoles 30-800 but is only
marginally consistent with a constant running beyond this range.  For this analysis, we have
implemented a $\sim$40$\times$ faster WMAP7 likelihood method which we have made publicly available.
\end{abstract}
\maketitle

\section{Introduction}

Under the assumption that cosmological perturbations were generated during an inflationary period from quantum fluctuations in a single scalar field, 
features in the cosmic microwave background (CMB) temperature
and polarization power spectra constrain features in the inflaton potential $V(\phi)$.

The usual slow-roll assumption of a  small and nearly constant dimensionless slope $(V'/V)^2$ and curvature
$(V''/V)$
of the inflaton potential leads to featureless power law initial curvature fluctuations.   
Breaking any of
these assumptions leads to features in the initial spectrum.
Indeed glitches in the observed temperature power spectrum of the CMB  \cite{Bennett:2003bz}
 have led to recent interest in exploring models with strong features in the potential
({\it e.g.}~\cite{Peiris:2003ff,Covi:2006ci,Hamann:2007pa,Mortonson:2009qv,Pahud:2008ae,Joy:2008qd,Dvorkin:2009ne,Hazra:2010ve}).  The significance at which such models are favored is difficult to address due to
the {\it a posteriori} manner in which they were chosen. 

Here we take a more model independent approach to constraining the
shape of the inflaton potential.   
We have recently shown that even in the presence of 
large local changes in the curvature of the inflaton potential that can explain the glitches, there is to excellent approximation only a single function of the inflaton potential that the observations constrain
\cite{Dvorkin:2009ne}.   Moreover, this function is approximately the same combination of
slope and curvature that enters into the calculation of the scalar tilt in the ordinary slow-roll approximation.  In this generalized
slow-roll (GSR) formalism this quantity need not be small or constant \cite{Stewart:2001cd,Choe:2004zg,Dodelson:2001sh}.   With it,
one can bypass a parameterization of the initial curvature power spectrum
({\it e.g.} \cite{Hu:2003vp,Bridle:2003sa,Leach:2005av,Sealfon:2005em,Paykari:2009ac,Nicholson:2009zj,Peiris:2009wp}) 
and the problem that not all spectra correspond to possible inflationary models.

In this Paper, we take a principal components approach \cite{Kadota:2005hv} to functional constraints on
the inflaton potential under the GSR formalism.   Principal components
constructed {\it a priori} from a noise model of the WMAP CMB measurements determine the theoretically best constrained deviations from a featureless potential 
before examining the actual data.
Constraints from the low order principal components thus efficiently encapsulate the expected information
content of the data and may be used to test a variety of inflationary models without
a reanalysis of the data.  

We begin in \S \ref{sec:GSR} with a brief review of the GSR formalism.  
In \S \ref{sec:PC} we develop the principal component analysis of the GSR source function
and apply it to the WMAP 7 year (WMAP7) data in \S \ref{sec:constraints}.   In \S \ref{sec:app} we consider
applications of these derived constraints on the inflaton potential and discuss
these results in \S \ref{sec:discussion}.  In the Appendix, we present the 
fast likelihood approach to the WMAP7 data employed in these analyses.

\section{Generalized Slow Roll}
\label{sec:GSR}

Given a specific model for the inflaton potential, the initial curvature fluctuation spectrum
can always be numerically computed 
by solving the linearized scalar field equation.  The slow
roll approaches, however, provide model independent mappings from the inflaton potential
to the curvature spectrum provided that the requisite approximations hold.  

The generalized slow roll (GSR) formalism was originally developed to consider cases where the usual slow roll
parameters $\epsilon_H \equiv (\dot \phi/H)^2/2$ and $\eta_H\equiv -\ddot \phi/H\dot \phi$ 
are small but $\eta_H$ is not necessarily constant
\cite{Stewart:2001cd,Choe:2004zg,Dodelson:2001sh}. 
Here $\phi$ is the inflaton field and overdots represent derivatives with respect to coordinate
time $t$.
In a previous paper \cite{Dvorkin:2009ne}, we
have shown that a variant of GSR works well for cases where $\eta_H$ becomes
of order unity for a fraction of an $e$-fold.  

In this variant of the GSR approximation, the curvature power spectrum is a non-linear functional 
\begin{eqnarray}
\label{eqn:ourGSR}
\ln \Delta_{\cal R}^2(k) &  \approx  & G(\ln \eta_{\rm min}) +  \int_{\eta_{\rm min}}^{\eta_{\rm max}} {d \eta \over \eta}  W(k\eta) G'(\ln \eta) \\
&&+ \ln\left[ 1 + {1 \over 2} \left(  \int_{\eta_{\rm min}}^{\eta_{\rm max}} {d \eta \over \eta} X(k\eta) G'(\ln \eta)  \right)^2\right]\,, \nonumber
\end{eqnarray}
of the function
\ba
G'(\ln\eta)={2\over 3}({f''\over f} - 3{f'\over f} - {f'^2\over f^2})
\ea
which is related to the inflaton potential through the background solution
 $f=2\pi \dot \phi a \eta/H$.  Primes here and below denote derivatives with respect to $\ln \eta$ and 
$\eta=\int_t^{t_{\rm end}} dt'/a$ 
is the conformal time to the end of inflation. 
We require $k_{\rm max} \eta_{\rm min} \ll 1$ and $k_{\rm min}\eta_{\rm max} \gg 1$ 
for the range in $k$ that we are calculating the spectrum.
 Integrating $G'$ gives
\begin{equation}
G(\ln\eta)=\ln({1\over f^2}) + {2\over 3}{f'\over f}\,.
\end{equation}

The window functions
\begin{eqnarray}
W(u) &=& {3 \sin(2 u) \over 2 u^3} - {3 \cos 2 u \over u^2} - {3 \sin(2 u)\over 2 u} \,, \nonumber\\
X(u) &=&  -{3 \cos(2u) \over 2 u^{3}} - {3 \sin(2 u) \over u^{2}} + {3 \cos(2 u) \over 2 u}  \nonumber\\&&
+
{3 \over 2 u^{3}}(1+ u^{2}) \,,
\end{eqnarray}
define the linear and nonlinear response of the curvature spectrum to $G'$ respectively.  
For the models we consider, the nonlinear response is small compared with the linear one.

The key property of Eq.~(\ref{eqn:ourGSR}) is that deviations from scale invariance in the curvature spectrum depend
only on a single function of time $G'$.  Moreover, to good approximation, this function
is related to the inflaton potential as  \cite{Dvorkin:2009ne}
\begin{equation}
G' \approx 3 ({V_{,\phi}\over V})^2 - 2{V_{,\phi\phi}\over V} \,,
\label{eqn:gprimeV}
\end{equation}
so long as $|\eta_H'| \gg |\eta_H|$ when $\eta_H$ is large, {\it i.e.}~that $\eta_H$ remains
large only for a fraction of an e-fold \cite{Dvorkin:2009ne}.
Finally, if the ordinary slow roll approximation where $\epsilon_H$ and $\eta_H$ are
both small and nearly constant holds, then  $G' $ 
may be evaluated at horizon
crossing $\eta \approx k^{-1}$ and taken out of the
integrals in Eq.~(\ref{eqn:ourGSR}).   As we shall see below, under this approximation 
$G'=1-n_s$.    By allowing $G'$ to be both time varying and potentially large,
we recover ordinary slow roll results where they apply but allow the data themselves
to test their validity.

\section{Principal components}
\label{sec:PC}

The GSR approximation allows us to go beyond specific  models of inflation in examining how the data constrain the inflaton potential.  
The data directly constrain the
function $G'$ and hence the derivatives of the inflaton potential through 
Eq.~(\ref{eqn:gprimeV}).

Given that $G'$ is related to the curvature spectrum $\Delta_{\cal R}^2$ by an integral
relation and the curvature spectrum itself is related to the observable  CMB 
power spectra by a line-of-sight integration, not all aspects of the function $G'$
are observable even with perfect data.   

We therefore seek a basis for an efficient representation of {\it observable} properties
of $G'$.  We begin in \S \ref{sec:basis} with a general description of a basis expansion
for $G'$ and its relationship to the usual normalization and tilt parameters.  We then turn in
\S \ref{sec:PC} to principal components (PCs) as the basis which best encapsulates
expected deviations from scale-free conditions \cite{Kadota:2005hv}.

\subsection{Basis Expansion}
\label{sec:basis}

In general, we seek to represent the function $G'$ as 
\begin{equation}
G'(\ln \eta)=\sum_{a=0}^N m_a S_a(\ln \eta) \,,
\label{eq:Gprime_reconstruction}
\end{equation}
where the basis functions $S_a$ for $a>0$ are assumed to have compact support in some
region between $\ln \eta_{\rm min}$ and $\ln \eta_{\rm max}$ corresponding to the
range probed by the data.   
We assume $S_0=1$ so that within this range $m_0$ represents a constant
tilt in the spectrum.

We seek  functions that obey the 
orthogonality and completeness relations
\begin{eqnarray}
{1 \over \Delta \ln \eta}  \sum_a S_a(\ln \eta) S_a(\ln \eta') &=& \delta(\ln\eta-\ln\eta')\,,\nonumber\\
{1 \over \Delta \ln \eta} \int d\ln \eta\, S_a(\ln\eta) S_b(\ln \eta)& = &
\delta_{ab} \,,
\label{eqn:orthocomplete}
\end{eqnarray}
where $\Delta \ln \eta = \ln \eta_{\rm max}- \ln\eta_{\rm min}$.
From these relations, the $m_a$ amplitudes are related to $G'$ as
\begin{equation}
m_a = {1 \over \Delta \ln \eta}\int d\ln \eta\,  S_a(\ln \eta) G'(\ln \eta) \,.
\label{eq:maprojection}
\end{equation}
Note that our normalization differs from that of 
Ref.~\cite{Kadota:2005hv} in that unit amplitude $m_a$ corresponds to unit variance
in $G'$ averaged over the whole
range in $\ln \eta$.

Substituting this model into the power spectrum expression (\ref{eqn:ourGSR}) yields
\ba
\ln\Delta_{\cal R}^2(k) &\approx& G(\ln\eta_{\rm min}) +m_0 C \nn\\
&&-m_0 \ln(k\eta_{\rm min})+  \sum_{a=1}^{N}m_aW_a(k)\\
&& + \ln\left[ 1 + {1 \over 2} \left(\sum_{a=0} ^N m_a X_a(k) \right)^2\right] \,,
\nonumber
\ea
where $C={7\over 3}-{\gamma_{\rm E}}-\ln\,2 \approx 1.06297$ with $\gamma_{\rm E}$ as
the Euler-Mascheroni constant, $X_0 = \pi /2$.   The $k$-space responses to the modes
are characterized by
\begin{eqnarray}
W_a(k) &=& \int_{\eta_{\rm min}}^{\eta_{\rm max}} d\ln \eta\, W(k\eta) S_a(\ln \eta) \,,\nonumber\\
X_a(k) &=& \int_{\eta_{\rm min}}^{\eta_{\rm max}} d\ln \eta\, X(k\eta) S_a(\ln \eta) \,, \nonumber
\end{eqnarray}
where $k\eta_{\rm min}\ll 1$ and $k\eta_{\rm max}\gg 1$.  Note that if $m_a=0$ for $a>0$ 
\be
\Delta_{\cal R}^2(k)  = e^{G(\ln \eta_{\rm min})+m_0 C}(1+{\pi^2 \over 8} m_0^2)(k\eta_{\rm min})^{-m_0} 
\ee
from which we infer that the model is a pure power law spectrum.
We can therefore choose instead to represent $G(\ln \eta_{\rm min})$
and $m_0$ by $n_s$ and $A_s$ bringing our parameterization of the power spectrum 
to
\ba
\label{eqn:PCspectrum}
\ln\Delta_{\cal R}^2 &=& \ln \left[ A_s \left( {k \over k_p} \right)^{n_s-1}\right] +  \sum_{a=1}^{N}m_aW_a(k) \\
&& + \ln\left[ 1 + {1 \over 2} \left( {\pi \over 2} (1-n_s) + \sum_{a=1} ^N m_a X_a(k) \right)^2\right]\,.
\nonumber
\ea
Note that this replacement ensures that the normalization and tilt parameters are
defined at a scale $k_p$ that is well-constrained by the data.  
Hence any
small and unobservable running of $G$ and $G'$ from $\eta_{\rm min}$ to $\eta \sim 1/k_p$ is absorbed into $A_s$ and $n_s$ respectively.  
Non zero values of $m_{a>0}$ required by the data thus represent a deviation from
purely scale-free initial conditions.  

We hereafter consider Eq.~(\ref{eqn:PCspectrum}) 
as the definition of the parameterized curvature power spectrum.
In practice we choose
$k_p=0.05$ Mpc$^{-1}$, and for reference note that $k \approx 0.02$ Mpc$^{-1}$ for
modes contributing to the well-measured first acoustic peak.

\begin{figure}[tbp]
\includegraphics[width=0.45\textwidth]{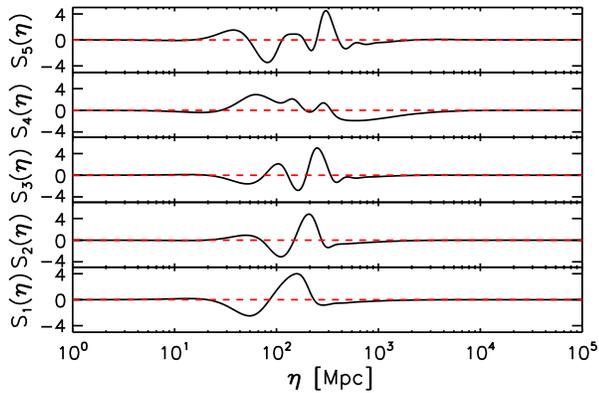}
\caption{\footnotesize The first 5 principal components (PCs) of $G'$  as a function of conformal time based on the WMAP7 specifications. 
The power law model with zero amplitude PCs is shown in red dashed lines.  The first
5 PCs represent a local expansion of $G'$ around $\eta \sim 10^{2}$ Mpc.
 }
\label{plot:eigenfunctions_WMAP7}
\end{figure}

\begin{figure}[tbp]
\includegraphics[width=0.45\textwidth]{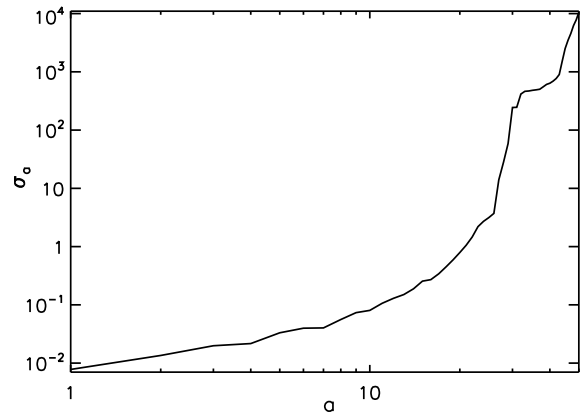}
 \caption{\footnotesize Predicted RMS error on the PC amplitudes as a function of mode number for WMAP7 data.  PCs are rank ordered from lowest to highest error with the first 5 describing 
 modes with better than $\sim 10\%$ constraints on $G'$.}
\label{plot:sigma_a_WMAP7}
\end{figure}

\subsection{Principal Component Basis}
\label{sec:PC}

We choose here to construct the basis functions $S_a$ from the principal components (PCs) of the projected WMAP7  covariance matrix for perturbations
in $G'$.  
To define the PCs, we begin with 
a fiducial flat $\Lambda$CDM model with a scale-free initial spectrum.  We take the baryon density to be
$\Omega_bh^2=0.02268$, cold dark matter density $\Omega_c h^2=0.1080$, cosmological constant $\Omega_\Lambda=0.7507$, optical depth
$\tau=0.089$, $A_s=2.41\times10^{-9}$, $n_s=0.96$ and $m_{a>0}=0$.  
This model corresponds to a constant $G'=0.04$. 

We then construct the PCs as 
 the theoretically best constrained non-constant deviations in
$G'$ around this fiducial model.
We start by adding a set of perturbations in $G'(\ln\eta_i) = 0.04 + p_i$
at 50 equally spaced intervals in $\ln \eta$ between
 $1<\eta/{\rm Mpc}<10^5$.  From this set, we define the 
 function
$G'(\ln \eta)$ by a cubic spline.  This sampling of 10 per decade or $\delta \ln \eta = 
0.23$ across $\Delta \ln \eta = 11.5$
 is sufficient to
capture the observable properties of $G'$ barring unphysical models with both
high frequency and high amplitude perturbations.  The spline ensures a smooth interpolation between
the samples.

\begin{figure}[tbp]
\includegraphics[width=0.45\textwidth]{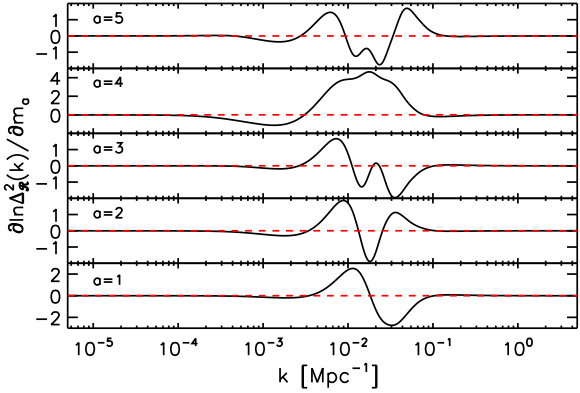}
\caption{\footnotesize Sensitivity of the curvature power spectrum to the first 5 PC parameters.
Low order PCs mainly change the power spectrum at wavenumber in the decade
around $k \sim 10^{-2}$ Mpc$^{-1}$.  Red dashed line represents the zero PC amplitude
fiducial power law model.}
\label{plot:dlnPofkdma_WMAP7}
\end{figure}

\begin{figure}[tbp]
\includegraphics[width=0.45\textwidth]{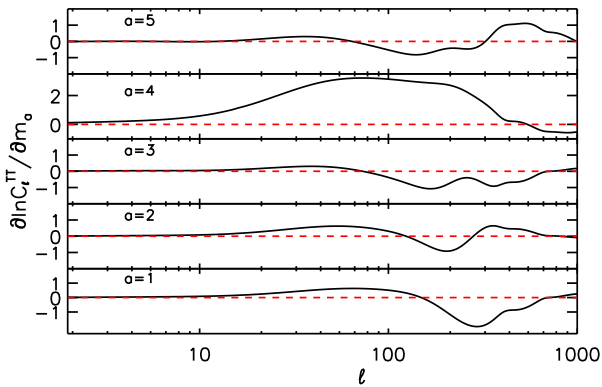}
\caption{\footnotesize Sensitivity of the temperature power spectrum to the first 5 PC parameters.
Low order PCs represent slowly varying features around the first peak $\ell \sim 200$.
Red dashed line represents the zero PC amplitude
fiducial power law model.}
\label{plot:dlnCldma_WMAP7}
\end{figure}

To these parameters $p_i$ we add the cosmological parameters 
\begin{equation}
p_{\mu}=\{p_{i},A_{s},n_{s},\tau,\Omega_bh^2,\Omega_{c}h^2,\theta\}\,,
\label{eq:parameters}
\end{equation}
where the angular extent of the sound horizon $\theta$ takes the place of $\Omega_\Lambda$,
 and construct the Fisher matrix
\be
F_{\mu\nu}=\sum_{\ell=2}^{\ell_{\rm max}}\sum_{XY,\tilde X\tilde Y}{\partial C_\ell^{XY}\over\partial p_\mu}{\bf C}^{-1}_{XY\tilde X\tilde Y}{\partial C_\ell^{\tilde X\tilde Y}\over\partial p_\nu},
\ee
where the $XY$ pairs run over the observable power spectra
$TT,EE,TE$. 
For the data covariance matrix, we take 
\begin{eqnarray}
{\bf C}_{XY\tilde X \tilde Y} &=& {1\over (2\ell+1)f_{\rm sky}} 
\Big[ 
(C_{\ell}^{X\tilde X} + N_\ell^{X\tilde X})
(C_{\ell}^{Y\tilde Y} + N_\ell^{Y\tilde Y}) \nonumber\\
&& 
+ (C_{\ell}^{X\tilde Y} +N_\ell^{X\tilde Y})
(C_{\ell}^{Y\tilde X} + N_\ell^{Y\tilde X}) \Big] \,,
\end{eqnarray}
where the WMAP7 noise power $N_\ell^{X\tilde X}=\delta_{X \tilde X} N_\ell^{XX}$ is inferred
from the temperature power spectrum errors from the LAMBDA
site \footnote{\url{http://lambda.gsfc.nasa.gov}} and the assumption that $N_\ell^{EE}= 2 N_\ell^{TT}$.
We set $\ell_{\rm max}=1200$ and $f_{\rm sky}=1$.

   We  then invert the Fisher matrix to form the covariance matrix
${\bf C} = {\bf F}^{-1}$.
Next we take the sub-block $C_{ij}$ and decompose it with the orthonormal matrix
$S_{ja}$, 
\begin{equation}
C_{ij} = { \Delta \ln \eta \over \delta \ln \eta} 
\sum_{a}S_{ia}\sigma_{a}^{2}S_{ja}
\end{equation}
rank ordered from lowest to highest $\sigma_a$.
Each eigenvector defines a discrete sampling of the basis function $S_a$ 
via
\begin{equation}
S_a(\ln\eta_i) =  \sqrt{ \Delta\ln \eta \over \delta \ln \eta} S_{ia} \,,
\end{equation}
with the normalization of Eq.~(\ref{eqn:orthocomplete}). 
 The full functions $S_a(\ln\eta)$ are
defined by taking a cubic spline through the samples.  
The first 5 PC components
are shown in Fig.~\ref{plot:eigenfunctions_WMAP7}.

In the Fisher approximation $\sigma_a$ represents the WMAP7 
expected errors for $m_a$
\begin{equation}
\langle m_a m_b \rangle  = \delta_{ab} \sigma_a^2 \,,
\end{equation}
for a zero mean fiducial model $\langle m_a \rangle =0$.

\begin{figure}[tbp]
\includegraphics[width=0.45\textwidth]{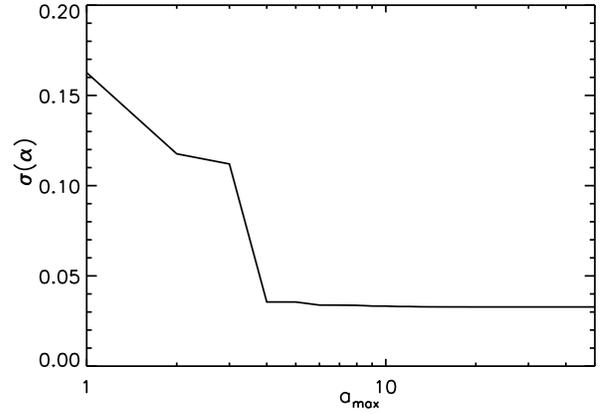}
\caption{\footnotesize Predicted RMS errors on running of tilt $\alpha$ as a function of the
maximum number of PC components included.  
Note that the errors cease to improve after the first 4 components and most of the
improvement comes from the 4th component.}
\label{plot:sigmaalpha_running_model_1d01d5_splined}
\end{figure}

These errors are shown in 
 Fig. \ref{plot:sigma_a_WMAP7}.   Note that the noise rises by a factor of 4 across the first
5 components and then increases to order unity by the 20th component.
Given that the peak amplitudes of $S_a$ lie in the $\sim 4$ range, a $\sim 2.5\%$ 
error corresponds to a $\sim 10\%$ peak variation in $G'$ or the effective tilt.  
By keeping the first 5 PCs, we retain all of the constraints on deviations  in $G'$ in the 
10\% range.  In this sense, these 5 modes represent the 
 best set for examining deviations from a scale-free power law model.
Note that no actual WMAP7 data goes into the construction and so that the modes
are chosen {\it a priori} rather than {\it a posteriori}.

For the first 5 PC components,
the $S_a$ basis functions are centered near
$\eta \approx 10^2$ Mpc and reflect the strong WMAP sensitivity to the first peak at $\ell 
\approx 200$ or $k \sim 0.02$ Mpc$^{-1}$ in the fiducial model.  The first 5 components resemble a local decomposition of $G'$ in
the decade surrounding this scale.   This fact can be seen more directly by examining
the sensitivity  of the curvature and temperature power spectra to the 5 PC amplitudes
(see  Figs. \ref{plot:dlnPofkdma_WMAP7} and \ref{plot:dlnCldma_WMAP7}, respectively).

As an example of the utility of retaining only the first 5 PCs, consider a linear model
for $G'$
\begin{equation}\label{eq:model_running}
G'(\ln\eta) = 1- n_s  + \alpha \ln(\eta/\eta_0) \,.
\end{equation}
The local slope in the power spectrum in the GSR approximation is
\begin{equation}
{d\ln \Delta^2_{\cal R} \over d\ln k}  = n_s-1 + \alpha\ln (k\eta_0) \,,
\end{equation}
and so $\alpha = dn_s/d\ln k$ and is equivalent to the running of the tilt.  
This linear model can be projected onto the first 5 PCs.
The variance of $\alpha$ is then given by
\begin{equation}
{1 \over \sigma^2(\alpha)} = \sum_{a=1}^{a_{\rm max}} {1 \over \sigma^2_a} 
\left( {\partial m_a \over \partial \alpha} \right)^2   \,,
\end{equation}
where
\begin{equation}
{\partial m_a \over \partial \alpha} ={1 \over \Delta \ln \eta} \int d\ln\eta \, S_a (\ln \eta) 
\ln(\eta/\eta_0)\,.
\end{equation}
By virtue of the marginalization of $n_s$ in the construction of the PCs, the $S_a$ functions
have nearly zero mean and $\partial m_a/\partial \alpha$ does not depend on the scale $\eta_0$ where the effective tilt is defined as $G'(\ln \eta_0)= 1-n_s$.

Fig. \ref{plot:sigmaalpha_running_model_1d01d5_splined} shows the predicted RMS error on $\alpha$ as a function of the maximum PC mode retained.
With $5$ PCs, $\sigma(\alpha)=0.0355$, while the fully-saturated value of the error with $50$ PCs is $\sigma(\alpha)=0.0327$.  This should be compared with the projected error using $\alpha$
itself as a Fisher matrix parameter, $\sigma(\alpha)=0.0328$.  These results verify the completeness of the 50 PC basis as well as show that most of the information for $|\alpha|\ll1$ is expected to come from the first 5 PCs.

In fact most of the information comes from a single mode, the 4th.  
This mode
corresponds to a local variation in the effective tilt $G'$ with a null near $300$Mpc
(see  Fig.~\ref{plot:eigenfunctions_WMAP7}),  or
in the power spectrum  near $\sim 0.003$  
Mpc$^{-1}$  (see Fig.~\ref{plot:dlnPofkdma_WMAP7}), and an extent spanning 1-2 decades.   Even though the 1st mode has smaller
overall errors and better constrains peak variations in $G'$, it is not the most effective
mode for constraining running of the tilt given its extremely local form.  

This caveat applies more generally.   A given model may have large deviations in 
$G'$ in a region where the data does not best constrain $G'$.  In this case the first 5 PCs no longer represent a complete or efficient representation.
We return to this point in \S \ref{sec:app}.

\begin{figure*}[btp]
\includegraphics[width=5in]{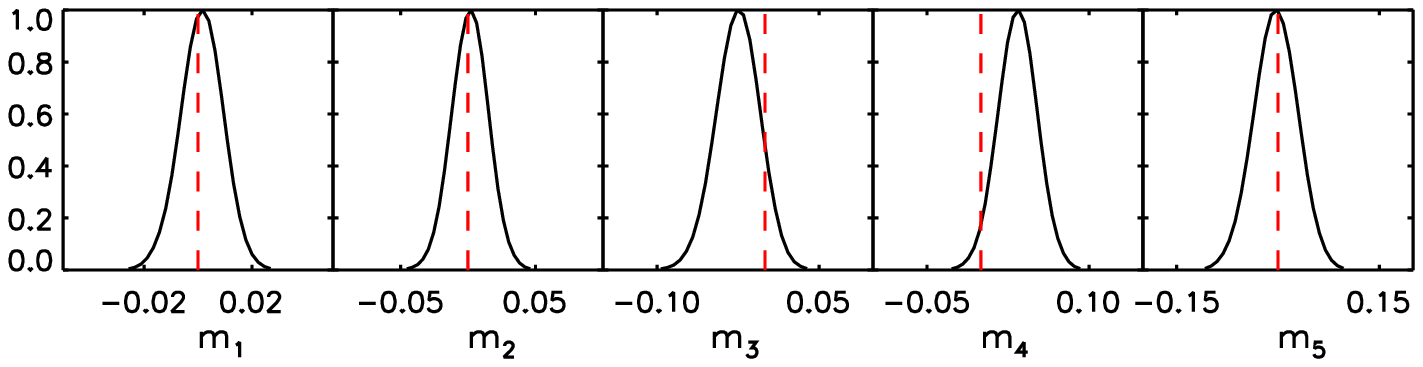}
\includegraphics[width=5in]{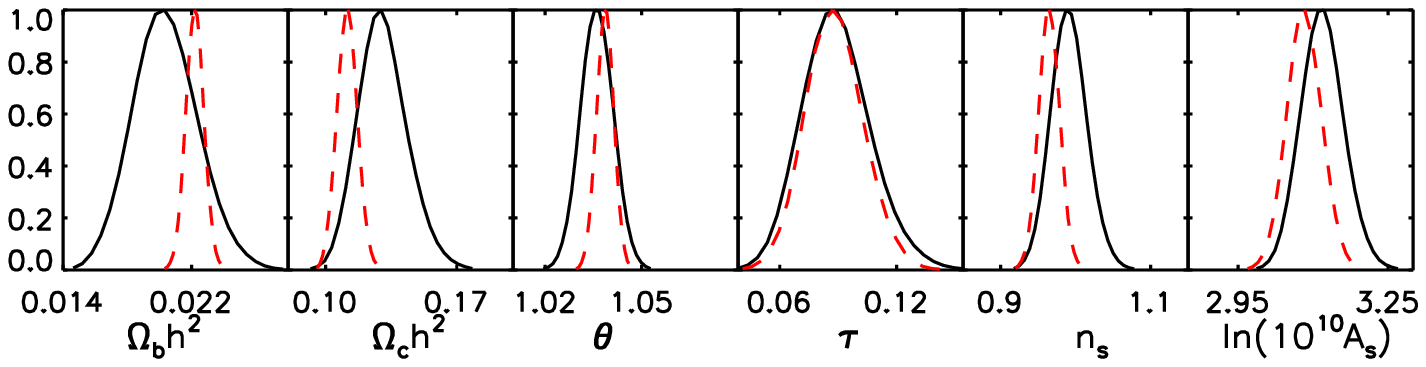}
\caption{\footnotesize Posterior probability distributions of the cosmological and 5 PC parameters
using WMAP7 data.   The red dashed line represents the 
 power law results where the first 5 PC parameters are held fixed to $m_{a}=0$.}
\label{plot:WMAP7_mas_5PCs_fast}
\end{figure*}

\subsection{MCMC}

We use a  Markov Chain Monte Carlo (MCMC) likelihood
analysis to determine joint constraints on the first 5 PC amplitudes and 
cosmological parameters
\begin{equation}
p_{\mu}=\{m_1,\ldots,m_5,A_{s},n_{s},\tau,\Omega_bh^2,\Omega_{c}h^2,\theta\}\,.
\label{eq:parameters}
\end{equation}
On top of this basic set we also examine the impact of spatial curvature $\Omega_K$ and
tensor-scalar ratio $r$ and the amplitude of a Sunyaev-Zel'dovich contaminant $A_{SZ}$ 
on a case-by-case basis.  

  The MCMC algorithm samples
the parameter space  evaluating the
likelihood ${\cal L}({\bf x}|{\bf p})$ of the data  ${\bf x}$  given each proposed parameter set ${\bf p}$
({\it e.g.}\ see~\cite{Christensen:2001gj,Kosowsky:2002zt}).  
The posterior distribution is obtained using Bayes' Theorem,
\begin{equation}
{\cal P}({\bf p}|{\bf x})=
\frac{{\cal L}({\bf x}|{\bf p}){\cal P}({\bf p})}{\int d\bm{\theta}~
{\cal L}({\bf x}|{\bf p}){\cal P}({\bf p})},
\label{eq:bayes}
\end{equation}
where ${\cal P}({\bf p})$ is the
prior probability density.  We place non-informative tophat priors on all parameters in
Eq.~(\ref{eq:parameters}).
For example, for the PC amplitudes we take ${\cal P}(m_{a>0})=1$ for $ -1 \le m_{a>0} \le 1$ 
and $0$ otherwise.

The MCMC algorithm generates random draws from the
posterior distribution that are fair samples of the likelihood surface.
We test convergence of the samples to a stationary distribution that
approximates the joint posterior density ${\cal P}({\bf p}|{\bf x})$ 
by applying a conservative Gelman-Rubin criterion \cite{gelman/rubin}
of $R-1< 0.01$ across four chains.
We use the
code CosmoMC \cite{Lewis:2002ah} for the MCMC analysis
\footnote{\url{http://cosmologist.info/cosmomc}}.

For the WMAP7 data \cite{Larson:2010gs}, we optimize the likelihood code available at
the LAMBDA web site as detailed in the Appendix.  
 The net improvement in speed on an 8-core
desktop processor is a factor of $\sim 40$ which should enable future studies with 20 or more PCs or other initial power spectrum parameters.

\begin{figure}[tbp]
\includegraphics[width=0.45\textwidth]{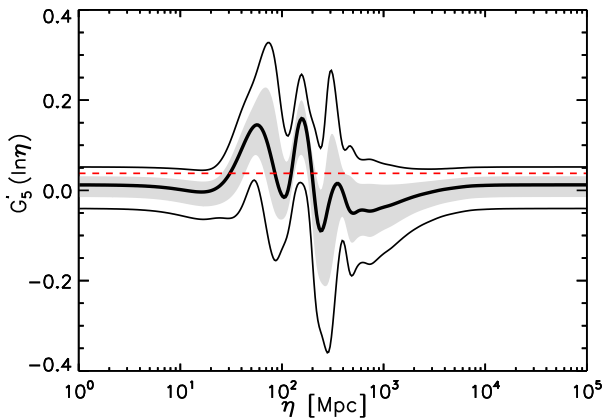}
\caption{\footnotesize The 5 PC filtered $G_5^{\prime}$ posterior using WMAP7 data. The shaded region encloses the $68\%$ CL region and the upper and lower curves show the upper and lower $95\%$ CL limits. The maximum likelihood (ML) $G^{\prime}_5$ is shown as the thick central curve, and the power law ML model is shown in red dashed lines. Structure in this representation mainly reflects
the form of the PC modes and is dominated by the modes with the largest
uncertainties.}
\label{plot:Gprime_posterior_5PCs_WMAP7_splined}
\end{figure}

\section{Constraints}
\label{sec:constraints}

In this section we present the constraints on the 5 best measured principal components of the GSR source function $G'$ and implicitly the inflaton potential $V(\phi)$.   We first examine constraints using
the WMAP7 data only and then joint with a variety of cosmological constraints  to remove residual parameter degeneracies.

\begin{figure}[tbp]
\includegraphics[width=0.45\textwidth]{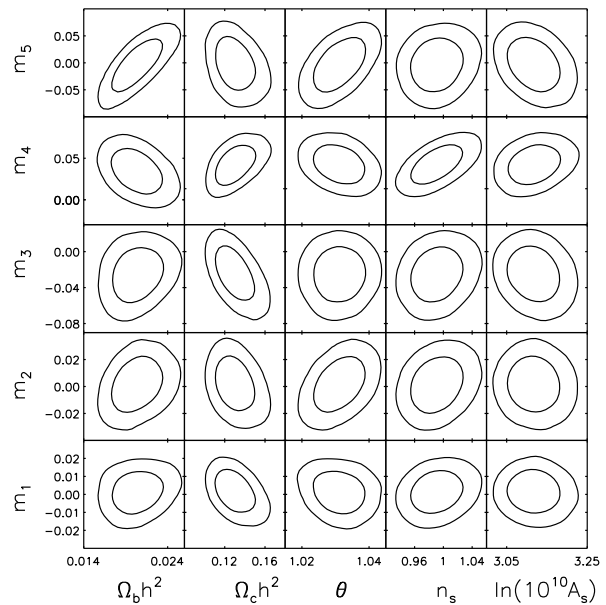}
\caption{\footnotesize Joint probability distributions of the principal component amplitudes and the cosmological parameters from MCMC analysis of  WMAP7 data (68\% and 95\% CL contours).}
\label{plot:2D_posteriors_m1_to_m5_WMAP7_1d01d5_splined}
\end{figure}

\subsection{WMAP}

We begin by considering the WMAP7 data in a flat $\Lambda$CDM cosmological context.
 We show the  probability distributions of the first 5 PCs and cosmological 
 parameters in Fig. \ref{plot:WMAP7_mas_5PCs_fast}.
Table \ref{table:WMAP7_5PCs} shows the mean, standard deviation of the posterior
probabilities as well as the maximum likelihood parameter values for the power
law models vs. the 5PC models.

For visualization purposes we show in Fig. \ref{plot:Gprime_posterior_5PCs_WMAP7_splined} the functional posterior probability of
\begin{equation}
G_5^{\prime}(\eta) \equiv 1-n_s + \sum_{a=1}^5 m_a S_a(\eta)\,,
\end{equation}
which should be interpreted as $G'$ filtered through the first 5 PCs and {\it not} a reconstruction
of $G'$ itself ({\it cf.}~Fig.~\ref{plot:Gprime_posterior_5PCs_WMAP7_1d01d5_splined_alldata_flat} below).

All 5 PCs are tightly constrained, with errors that are comparable to the Fisher
matrix projection, nearly Gaussian posteriors and little covariance with each other. 
The correlation coefficients between two different $m_a$'s, $|C_{m_a m_b}/\sigma({m_a})\sigma({m_b})|<0.2$.   The first component $m_1$ in particular is consistent with zero
and places the tightest constraints of $\lesssim 3\%$
 local variations in $G'$ around $\eta \approx 10^2$ Mpc.
Interestingly, the power law prediction of $m_4=0$ lies in the tails of the posterior with
as extreme or more values disfavored at $94.8 \%$CL.   With the Gaussian approximation $m_4=0$ is
$1.9\sigma$ from the mean and disfavored at $94.5\%$CL.

  In Fig.~\ref{plot:2D_posteriors_m1_to_m5_WMAP7_1d01d5_splined}, we show joint posteriors of the PCs with other parameters.  Notably, for the anomalous $m_4$ component there is a degeneracy with $\Omega_c h^2$, $\Omega_b h^2$ and 
  $n_s$ which is also reflected in the broadening of the cosmological parameter 
  posteriors in Fig.~\ref{plot:WMAP7_mas_5PCs_fast} and the shift in means and 
  maximum likelihood values in Table \ref{table:WMAP7_5PCs}.    The maximum likelihood (ML)
  model found by the chain is an improvement over the power law case
  of $2\Delta \ln {\cal L} \approx 5$ which is marginal considering the addition of 5 parameters.  
  
 The intriguing aspect of the ML model, like the $m_a$ posteriors, is that the improvement
  is concentrated in the $m_4$ component. Note that the finite $m_4$ component
  allows $n_s=1$ to be a good fit to the data implying that the data can be
  marginally better fit by a local deviation from scale invariance rather than tilt.
    
\begin{figure}[tbp]
\includegraphics[width=0.45\textwidth]{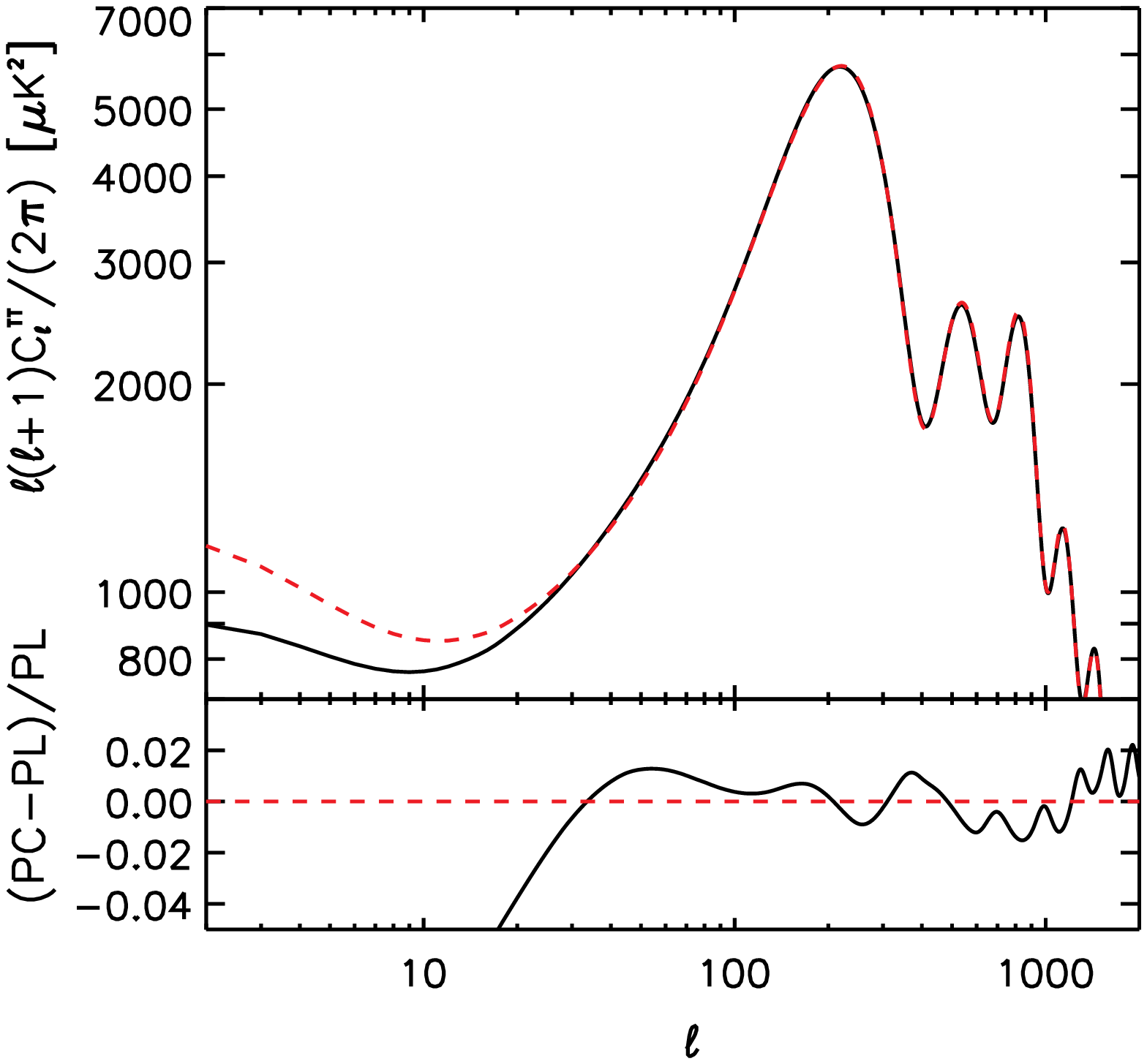}
\includegraphics[width=0.45\textwidth]{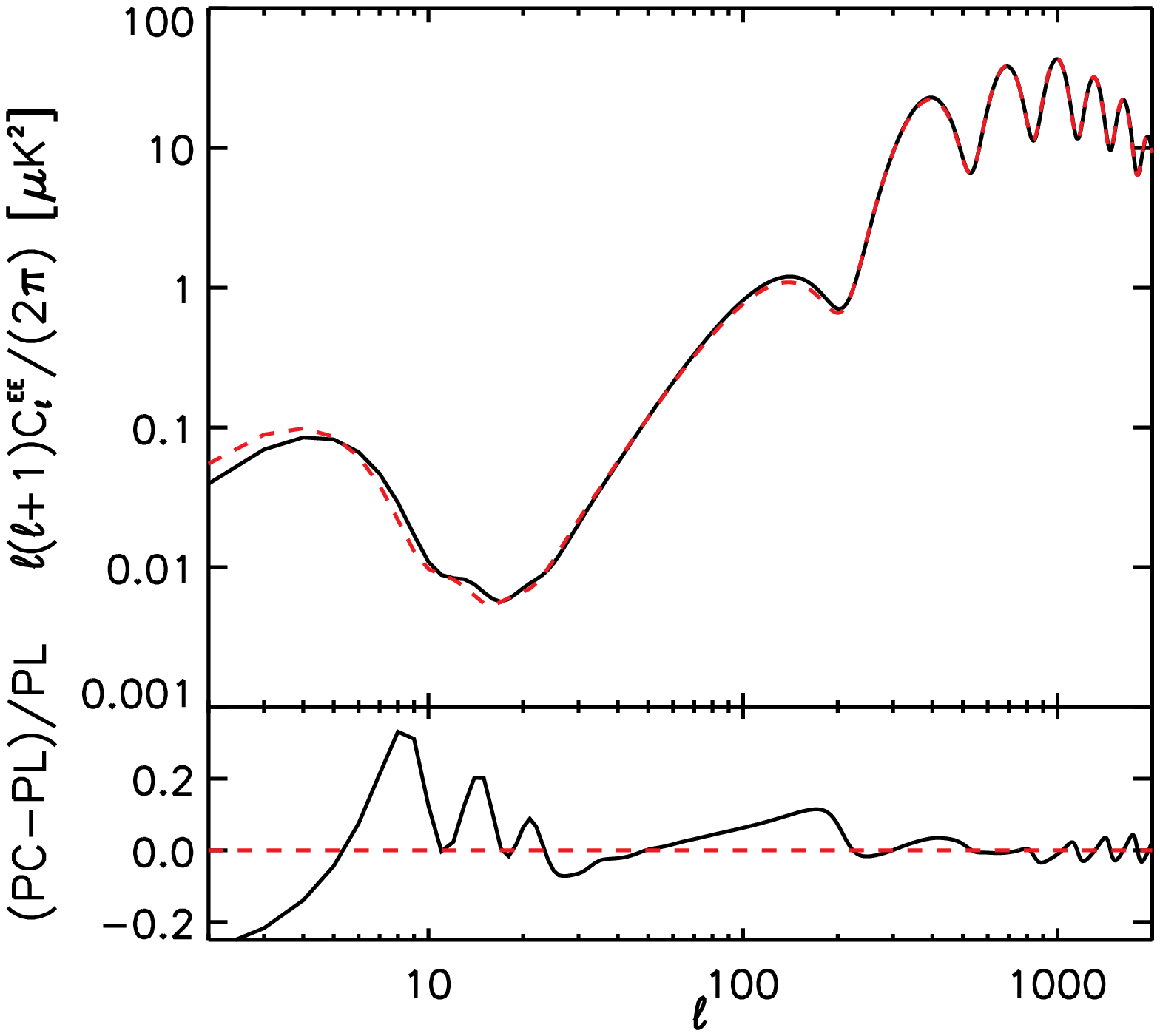}
\caption{\footnotesize Power spectra of the $5$ PCs maximum likelihood model (black curve) compared to power law (PL) maximum likelihood model (red dashed curve) in the upper panel, and the  difference (PC-PL)/PL  in the lower panel.  Top: temperature power spectrum.  Bottom: polarization power spectrum.}
\label{plot:5PCs_WMAP7_ML_TT}
\end{figure}

   An examination of the ML model
  helps illuminate the degeneracies with cosmological parameters.
  Fig. \ref{plot:5PCs_WMAP7_ML_TT} (top)  shows the temperature power spectra of the $5$ PC ML model compared to the power law ML model (upper panel), and the fractional difference between the two (lower panel). 
  Note that in the well constrained $\ell \sim 30-800$ regime the two spectra agree to
  $\sim 1\%$ or better. 
     This accidental degeneracy is not preserved beyond 
  $\ell=1000$.  Furthermore the  $E$-mode polarization power spectra shown in
  Fig. \ref{plot:5PCs_WMAP7_ML_TT} (bottom) reveal substantially larger fractional deviations
  of up to $\sim 10\%$ that break the degeneracy in the temperature 
  spectra.

\begin{figure}[tbp]
\includegraphics[width=0.45\textwidth]{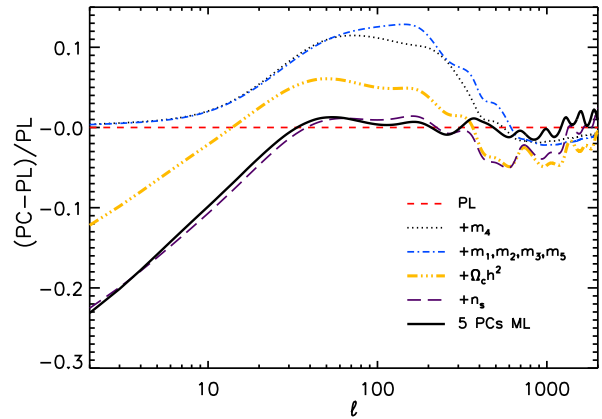}
\caption{\footnotesize Decomposition of the fractional difference between 
the PC and power law (PL) maximum likelihood models (ML) shown in Fig.~\ref{plot:5PCs_WMAP7_ML_TT} into contributions from specific parameters.   Curves show the
cumulative effect of adjusting the PL ML parameters to the PC ML values.
The main effects come from the change in $m_4$, $\Omega_ch^2$, and $n_s$.}
\label{plot:PL_to_5PCs_ML}
\end{figure}

Indeed  the main improvement of the PC model relative to the PL model actually comes from the $\ell \ge 24$ polarization cross correlation part of the likelihood (MASTER $TETE$), where $2\Delta \ln {\cal L}=3.5$, with half of this contribution coming from $\ell<200$.  The low-$\ell$ temperature part of the likelihood has an improvement of $2\Delta \ln {\cal L}=1.8$ relative to the PL model, and there is a smaller improvement coming from the high-$\ell$ (MASTER $TT$) part with $2\Delta \ln {\cal L}=1.1$. Finally, the PC model is worse than the PL model by $2\Delta \ln {\cal L}=-1.3$ for the low $\ell<24$ polarization.

  The intermediate $\ell$ temperature degeneracy exhibited by the two models is further explored in
  Fig. \ref{plot:PL_to_5PCs_ML}.  We show the impact of the cumulative parameter variations between the PL and PC ML models.  The addition of $m_4$ carries almost all
  of the impact of the PCs and is mainly compensated by variations in $\Omega_c h^2$ to adjust the relative heights of the peaks and $n_s$ to tilt the spectrum to small
  scales.
    The change in $\Omega_c h^2$ also changes the physical size of the sound horizon
    which must be compensated by a change in the distance to recombination reflected
    in a lower value for $H_0$ and $\Omega_\Lambda$ to leave the angular scale $\theta$
    compatible with the data.

These results are robust to marginalizing $\Omega_K$ given any reasonable prior on $H_0$,
an SZ component through $A_{SZ}$, or tensors within the $B$-mode measured limits of the BICEP experiment $r = 0.03^{+0.31}_{-0.26}$ \cite{Chiang:2009xsa}.  Through an explicit MCMC analysis of these separate cases, we have verified that
 the shift in the means and change in the errors for all 5 PCs are much smaller than
 $1\sigma$.  The largest effect is from marginalizing tensors where
 for example $m_4 = 0.042 \pm 0.20$.  A small improvement in the $B$-mode limits would eliminate
  this ambiguity entirely.

\begin{table*}[ht]
\centering
\begin{tabular}{|l|r@{}|c|c|r@{}|c|c|}
\hline
Parameters & & \multicolumn{2}{|c|}{Power Law (PL)} & & \multicolumn{2}{|c|}{Principal Components (PC)} \\
\cline{1-1} \cline{3-4} \cline{6-7} 
$100\Omega_b h^2$  & & $2.220\pm0.055$   & $2.217$  & & $2.040\pm0.196$   & $2.067$ \\
$\Omega_c h^2$     & & $0.1116\pm0.0053$ & $0.1130$ & & $0.1308\pm0.0127$ & $0.1284$ \\
$\theta$           & & $1.0386\pm0.0026$ & $1.0387$ & & $1.0361\pm0.0049$ & $1.0365$ \\
$\tau$             & & $0.088\pm0.014$   & $0.088$  & & $0.089\pm0.017$   & $0.088$ \\
$n_s$              & & $0.9650\pm0.0136$   & $0.9622$  & & $0.9916\pm0.0233$   & $0.9877$  \\
$\ln[10^{10}A_s]$  & & $3.083\pm0.034$   & $3.088$  & & $3.119\pm0.041$   & $3.105$ \\
$m_1$&& 0 & 0 &&$0.0014\pm0.0077$ & $0.0021$ \\ 
$m_2$&& 0 & 0 &&$0.0015\pm0.0132$ & $0.0068$ \\
$m_3$&& 0 & 0 &&$-0.0253\pm0.0197$ & $-0.0264$ \\
$m_4$&& 0& 0 &&$0.0339\pm0.0175$ & $0.0337$ \\
$m_5$&& 0 & 0 &&$-0.0033\pm0.0315$ & $0.0023$ \\
\cline{3-4} \cline{6-7} 
$H_0$              & & $70.13\pm2.38$      & $69.50$   & & $61.41\pm6.08$      & $62.18$  \\
$\Omega_\Lambda$ && $0.726\pm0.028$ & $0.720$ && $0.581\pm0.116$ & $0.614$ \\
\cline{1-1} \cline{3-4} \cline{6-7} 
$-2\ln {\cal L}$ & & \multicolumn{2}{|c|}{$7474.97$} & & \multicolumn{2}{|c|}{$7469.82$} \\
\cline{1-1} \cline{3-4} \cline{6-7} 
\end{tabular}
\caption{\footnotesize Means, standard deviations (left subdivision of columns) and maximum likelihood values (right subdivision of columns) with likelihood values for $\Lambda$CDM and the $5$ PCs model with WMAP7 data.}
\label{table:WMAP7_5PCs}
\end{table*}

\subsection{Joint Constraints}

The results of the previous section suggest that other data which measure the high-$\ell$ temperature spectrum,
 polarization spectrum, or pin down the cosmological parameters that control the distance
to recombination and baryon density 
can eliminate the remaining degeneracies and enable WMAP7 to better
constrain the inflaton potential.

We start with adding more CMB information from the QUAD experiment.  
QUAD helps mainly by reducing the $m_5-\Omega_bh^2$ degeneracy.  Interestingly, most of the impact comes from the polarization measurements rather than
the extended range of the temperature constraints.

Adding in non-CMB cosmological information helps even more, especially with $m_4$.    
We take the UNION2 supernovae dataset \footnote{\url{http://www.supernova.lbl.gov/Union}}, the SHOES $H_0= (74.2 \pm 3.6)$ km/s/Mpc measurement \cite{SHOES}
and a big bang nucleosynthesis constraint of $\Omega_b h^2 = 0.022 \pm 0.002$.
In a flat $\Lambda$CDM cosmology, the degeneracy between $m_4$ and $\Omega_c h^2$ is
nearly eliminated yielding $m_4 = 0.0191 \pm 0.0163$, {\it i.e.}~consistent with power law 
initial conditions.    Of the additional data it is the supernovae that drive this improvement
by disfavoring the low $\Omega_\Lambda$ values required by the increase in 
$\Omega_c h^2$.   

On the other hand, these improvements require an assumption that the dark energy
is a cosmological constant and the Universe is spatially flat.   For example if
$\Omega_K$ is marginalized, $m_4 = 0.0384 \pm 0.0197$.   The addition of spatial curvature allows the freedom to adjust 
the relative distance to the high-$z$ supernova and recombination.
A better measurement of
$H_0$ could resolve this degeneracy since the constraints on $\Omega_K$ are already  
limited by the SHOES data.

Table \ref{table:datasets_5PCs} summarizes these results for the constraints on the PC 
amplitudes.

\begin{table*}
\begin{center}
\begin{tabular}{|c|c|c|c|}
\hline
PC & +QUAD & +BBN$+$SN$+$$H_0$, flat & +BBN$+$SN$+$$H_0$, w/$\Omega_K$ \\
\hline
\hline
$m_1$ & $0.0000\pm0.0072$ & $0.0045\pm0.0071$ & $0.0027\pm0.0073$ \\
$m_2$ & $0.0033\pm0.0123$ & $0.0091\pm0.0121$ & $0.0045\pm0.0125$ \\
$m_3$ & $-0.0261\pm0.0184$ & $-0.0120\pm0.0166$ & $-0.0208\pm0.0178$ \\
$m_4$ & $0.0296\pm0.0168$ & $0.0191\pm0.0163$ & $0.0384\pm0.0197$ \\
$m_5$ & $0.0149\pm0.0250$ & $0.0187\pm0.0249$ & $0.0091\pm0.0256$ \\
\cline{1-4}
\end{tabular}
\caption{\footnotesize Means and standard deviations of the posterior probabilities of the PC amplitudes with different data sets added to the WMAP7 data. For supernovae (SN) we used the UNION$2$ dataset, and for $H_0$ the SHOES measurement.}
\label{table:datasets_5PCs}
\end{center}
\end{table*}

\section{Applications}
\label{sec:app}

The model independent results of the previous section can be used to test a wide variety of inflationary deviations from scale-free conditions.  Moreover 
given that the constraints on the PC amplitudes are uncorrelated and approximately Gaussian, these tests are straightforward to apply.

As the simplest example, take the running of the tilt model defined by $\alpha=dn_s/d\ln k$
through Eq.~(\ref{eq:model_running}).  Using Eq.~(\ref{eq:maprojection}), we
obtain
\begin{eqnarray}
&& m_1 = 0.048\alpha \,,\quad
m_2 = -0.079\alpha \,, \quad 
m_3 = 0.054 \alpha \,,\nonumber\\
&& m_4 = -0.576 \alpha \,,\quad 
m_5 = -0.034 \alpha \,. 
\label{eq:malpha}
\end{eqnarray}
We can then construct the effective $\chi^2$ statistic
\begin{equation}
\chi^2(\alpha) = \sum_{a=1}^5  \left[ { m_a(\alpha) - \bar m_a \over \sigma_a} \right]^2 \,.
\end{equation}
With the means and variances taken from Table \ref{table:WMAP7_5PCs} for WMAP7
we obtain $\alpha=-0.057\pm0.029$.  
When we take into account the covariance between the PC amplitudes, we obtain a $3\%$ shift in the mean with the same error: $\alpha=-0.059\pm0.029$.  Likewise we have verified
that using Eq.~(\ref{eq:malpha}) in a separate MCMC analysis gives consistent results
$\alpha = -0.058 \pm 0.030$.

This result should be compared with the direct analysis of the running of the tilt which
gives  $\alpha = -0.034 \pm 0.027$ consistent with the analysis from \cite{Larson:2010gs}.  
The mean is shifted from the PC derived mean by $\sim 0.8\sigma$ while the errors are $7\%$ higher.  

The small shift in the mean is driven mainly by the truncation to 5 PC components.
This is in spite of the Fisher expectation in \S \ref{sec:PC} that for an infinitesimal $\alpha$
the first 5 PC components contain nearly all the information.  A running of the tilt model
where $\alpha = -0.057$, which fits the intermediate $\ell \sim 30-800$ range well, 
implies a large change across the extended observable
range from $\eta \sim 20-5000$ Mpc of $|\delta G' |  \sim |\delta n_s|  \sim  0.3$.  In particular, it overpredicts the suppression
of the $\ell < 30$ temperature multipoles.   For the same amplitude and tilt 
at the first peak, the amplitude at the horizon is suppressed by
$\sim e^{(\alpha/2)\ln^2(100)} = 0.55$.  This suppression can only be partially compensated by red tilting the spectrum without over suppressing the high $\ell >800$ multipoles.
Note that cosmic variance completely dominates the uncertainties in the $\ell<30$ region and
decreases with the predicted signal, an effect that is not represented in the Fisher matrix.

   In other words, the $2\sigma$ preference for
finite $m_4$ is not completely consistent with a constant running of the tilt but rather points to a more local deviation from scale-free conditions.
When we eliminate this preference by adding in the additional SN, $H_0$ and BBN constraints the inferred limits on $\alpha$ from the first 5 PC amplitudes
becomes $\alpha = -0.033 \pm 0.027$.  Note however that the direct constraints on $\alpha$
also improve with the addition of these data sets to $\alpha = -0.013 \pm 0.021$.

\begin{figure}[tbp]
\includegraphics[width=0.45\textwidth]{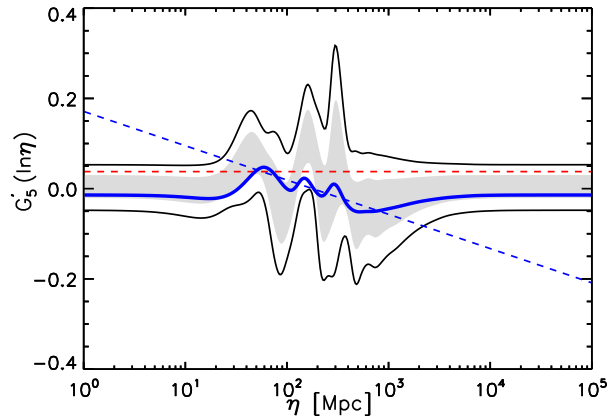}
\caption{\footnotesize The 5 PC filtered $G_5^{\prime}$ posterior using WMAP$7$ data and additional SN, $H_0$ and BBN constraints in a flat universe. The shaded region encloses the $68\%$ CL region and the upper and lower curves show the upper and lower $95\%$ CL limits. A model with running of the tilt $\alpha=-0.033$, the mean value given these constraints,  is shown as the thick solid blue curve and the ML PL model as the dashed red curve. 
The 
unfiltered $G'$ of the same $\alpha$ is shown in blue dashed lines for comparison (arbitrary offset).
Note that outside the range probed by the first 5 PCs the model deviations  continue to
grow linearly and oscillating features in $G'_5$ do not necessarily imply features in the underlying $G'$.}
\label{plot:Gprime_posterior_5PCs_WMAP7_1d01d5_splined_alldata_flat}
\end{figure}

In Fig. \ref{plot:Gprime_posterior_5PCs_WMAP7_1d01d5_splined_alldata_flat} we plot an $\alpha = -0.033$ model for the $5$ PCs filtered $G_5'$ against the posterior constraints from the WMAP$7$ data and additional SN, $H_0$ and BBN constraints in a flat universe.
We overplot the original unfiltered $G'$ for this $\alpha$ and note that even with
the reduced value the deviations become large outside of the region probed by the
first 5 PCs.

This example points out a caveat to the use of the first 5 PCs as general constraints
on models.  
For a model with features that are substantially larger in
a regime away from the well-constrained first acoustic peak, the first 5 PCs may not be the best
constraints in terms of signal-to-noise.   One can check whether this is the case by examining
the predicted $G'$ or by projecting
the model onto the full 50 PC space and checking for large amplitude components.
Indeed if the higher components are extremely large compared with the low components,
non-linear effects can break the orthogonality of PCs and lead to larger allowed variations in the low components when compensated by the high components.

\begin{figure}[tbp]
\includegraphics[width=0.45\textwidth]{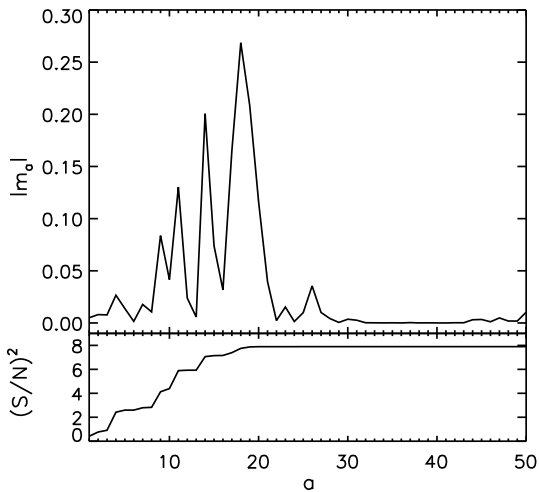}
\caption{\footnotesize Principal component amplitudes for the step function potential model 
\cite{Mortonson:2009qv} that
best fits the glitches in the temperature spectrum at  $\ell\sim20-40$ (upper panel), and 
projected cumulative 
$(S/N)^2 = \sum (m_a /\sigma_a)^2$ (lower panel).  Given the large values of $m_a$ in the
high order PC components, $\sim 20$ PCs are required to completely characterize this model.}
\label{plot:ma_wigglyML}
\end{figure}

As an example, the full 50 PC decomposition of the  the step function potential model from \cite{Mortonson:2009qv}
which fits the $\ell=20-40$ glitches in the temperature power spectrum is shown in Fig.~\ref{plot:ma_wigglyML}. Interestingly, $m_4=0.0266$ in the step model and has the highest amplitude of the first 5 components.  On the other hand,   a complete analysis based on signal-to-noise 
would require $\sim 20$ PC components.  By keeping only 5 components, the
improvement compared with the ML PL model is only $\Delta \chi^2 = -1.7$.  In other 
words, the step function model is certainly  allowed by our 5 PC constraint and even marginally
favored but the majority of the improvement is not captured by the truncated analysis.

Nevertheless,
when interpreted as an upper limit on deviations from scale-free conditions, the  5 PC approach
works as a general, albeit typically conservative, method to constrain a wide variety of possible deviations from 
a single analysis.   As the running of the tilt example shows, the results are remarkably
close to a direct analysis and differences can be used to expose the self-consistency of
the model inferences with independent parts of the
data.

\section{Discussion}
\label{sec:discussion}

We have employed a variant of the generalized slow roll approximation (GSR) that allows large amplitude deviations from scale-free conditions to place
functional constraints on the GSR source function and implicitly on the inflaton potential.  By employing a principal component (PC) decomposition, we isolated 
the 5 best functional constraints imposed by the WMAP7 data.
The analysis is greatly facilitated by our optimization of the WMAP7 likelihood
code which we have made publicly available \footnote{
\url{http://background.uchicago.edu/wmap_fast}}.

These 5 PCs provide incisive constraints on the inflaton potential around the
e-folds of inflation when the scales associated with the
 first acoustic peak were crossing the horizon.  Non-zero values for their amplitudes
 represent deviations from slow
 roll and power law initial spectra.  The first component implies that deviations are less than $\sim 3\%$ near 
 $\eta \sim 10^2$ Mpc and the first 5 represent constraints around that scale
 at better than the 10\% level.  The result is 5 nearly
 independent Gaussian constraints that can be applied to any inflationary model
 where this level of deviation is expected.  We have also made the eigenfunctions, which are required to project a given model onto the PC amplitudes, publicly available. These limits are robust to the inclusion of
 tensor contributions allowed by current $B$-mode limits, spatial curvature and
  Sunyaev-Zel'dovich contamination from unresolved clusters.
 
 Interestingly, for the 4th principal component the null prediction of scale-free initial conditions
 is disfavored at the 95\% CL.    However, given the 5 added parameters,
 this result does not rule out a power law initial spectrum at a significant level.  
 Moreover, the relatively large deviations implied by this anomalous mode 
 are allowed only through correspondingly large variations in the cosmological
 parameters, mainly the cold dark matter and its effect on the sound horizon and by inference the distance to recombination.  
 
  Further information from the CMB polarization
 and high-$\ell$ temperature power spectrum can break this degeneracy.  
 The QUAD polarization data already have some impact on the constraint and the Planck satellite
 should definitively resolve this issue.    External data also can break this degeneracy.
 In particular in a flat $\Lambda$CDM cosmology, the distance to high redshift supernovae
 reduce the preference for finite $m_4$ from 1.9$\sigma$ to 1.2$\sigma$.  However this
 improvement disappears if spatial curvature is marginalized.
 
 This anomalous 4th PC resembles a {\it local} running of the tilt
 around scales of $\eta \sim 300$Mpc.    Direct analysis of a {\it global} constant
 running of the tilt shows that this preference is mainly local, {\it i.e.}~the low and
 high multipoles 
 prefer a different and smaller running than the intermediate
  multipoles that the first 5 PCs probe.
 The running of the tilt example illustrates the use of the PC constraints
 both as a  technique to constrain inflationary parameters arising from different models
  with a general analysis
 and as a method for examining what aspects of the data drive the constraints.
 
The running of the tilt example also illustrates that for models where  deviations
from scale-free conditions become much larger than $\sim 10\%$ away from the well-constrained region
of the acoustic peaks, more principal components are required to ensure a 
complete and incisive description.   We intend to examine these issues in a future
work.

\smallskip
{\it Acknowledgments:} We thank Michael Mortonson, Eric Switzer and Amol Upadhye for useful conversations.
This work was supported by the KICP under
NSF contract PHY-0114422.   WH was additionally supported by DOE contract DE-FG02-90ER-40560 and the Packard Foundation.
\appendix

\begin{figure}[tbp]
\includegraphics[width=0.5\textwidth]{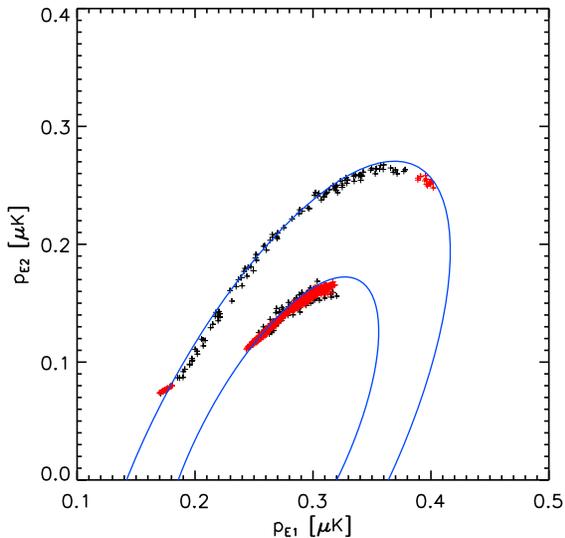}
\caption{\footnotesize Comparison of the low-$\ell$ polarization pixel likelihood $-2 \ln {\cal L}_{\rm pol}^{\ell<24}$ and the approximate fit 
as a function of $E$-mode polarization amplitude in
two multipole bands ${p}_{E 1}$ ($\ell = 4-6$), ${p}_{E 2}$ ($\ell=8$).   Models from a power law chain (red crosses) and from a $5$ PCs chain (black crosses) whose likelihood relative to the minimum 1645.84
are (within $\pm 0.1$) of $2.29$ (68.27\% CL) and $6.18$ (95.45\% CL) are shown
with the contours from the fit overplotted (blue curves).}
\label{plot:contour_2bands_pE456_pE8}
\end{figure}

\section{Fast WMAP Likelihood Evaluation}

In this Appendix, we describe the optimization of the WMAP likelihood code and fast
approximate techniques for describing the low-$\ell$ polarization information.
 Changes in the initial power spectrum do not require recomputation of 
the radiation transfer function and are so-called fast parameters for CosmoMC.  Hence
the WMAP7 likelihood computation is the main bottleneck for the MCMC analysis.

We first OpenMP 
parallelize the likelihood code and remove bottlenecks in the computation of
the temperature and high-$\ell$ polarization likelihood.   We obtain a $\sim 2.6 N_{\rm core}$
speedup of those parts of the likelihood where $N_{\rm core}$ is the number of cores in a shared memory machine.     These changes exactly preserve the accuracy of the
likelihood evaluation.

\begin{figure}[tbp]
\includegraphics[width=0.45\textwidth]{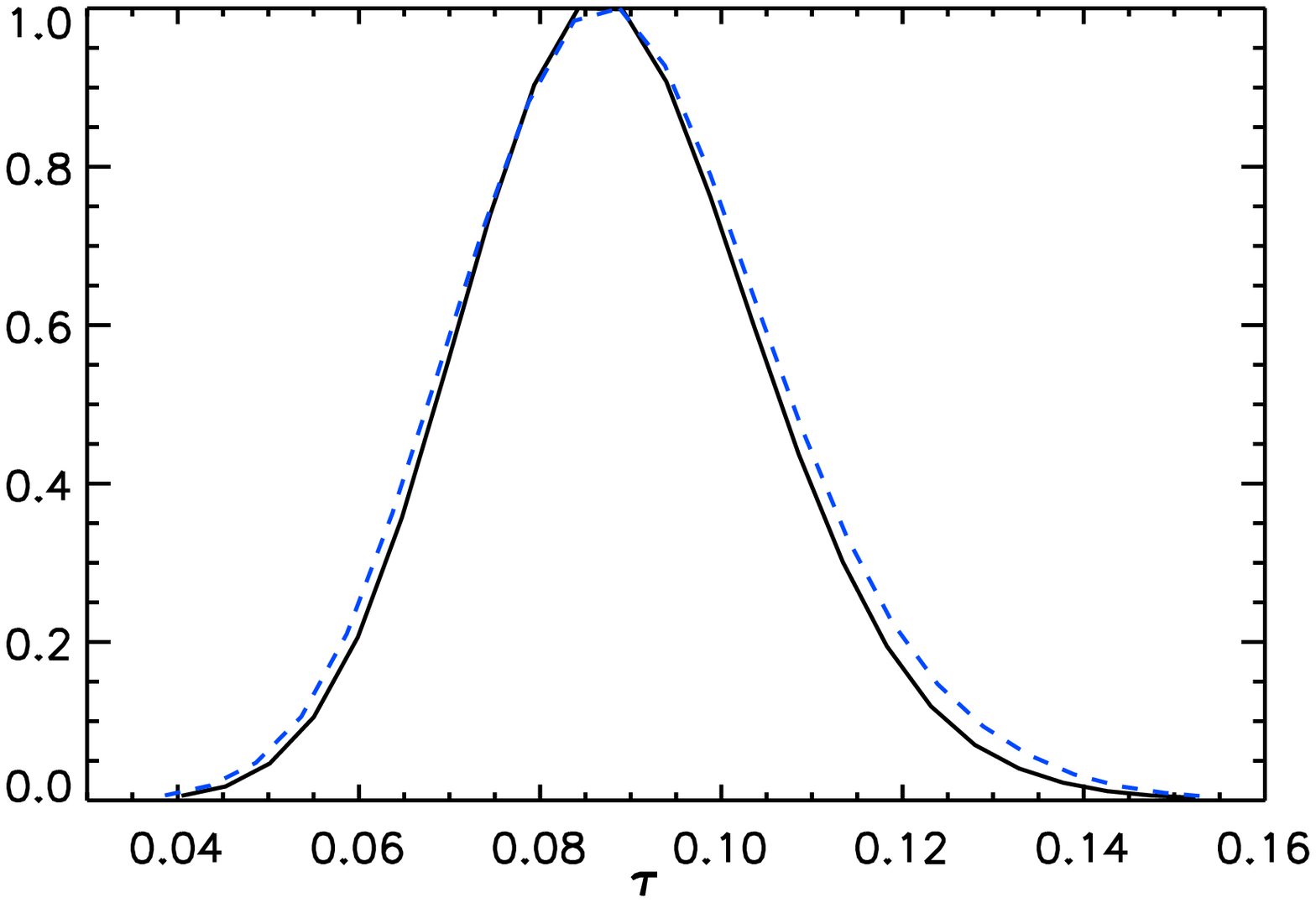}
\includegraphics[width=0.45\textwidth]{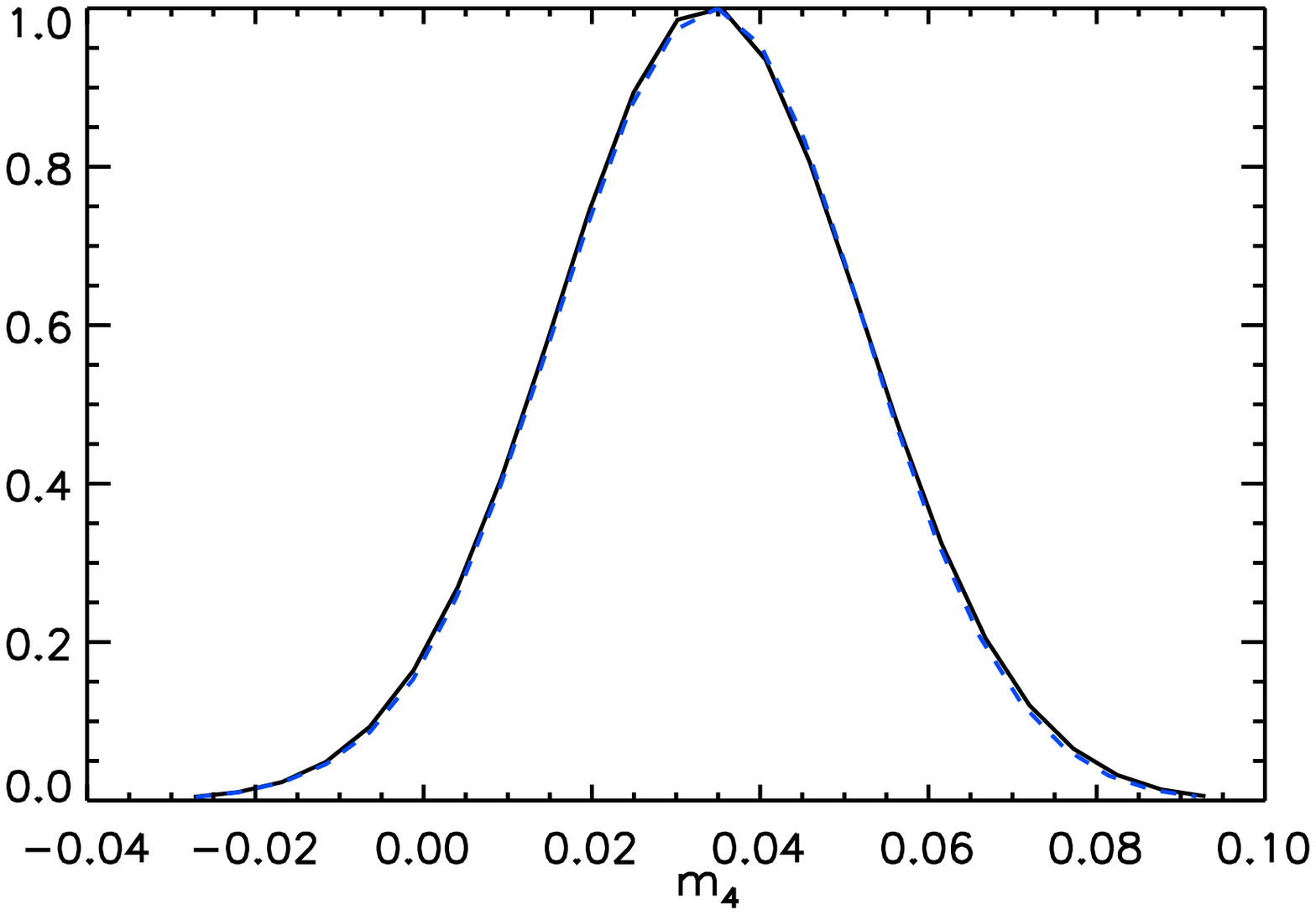}
\caption{\footnotesize Posterior probability distribution of the optical depth $\tau$ and the fourth PC amplitude using the exact likelihood (black curves) and the approximation (blue dashed curves) with WMAP data only.}
\label{plot:tau_m4_fullL_1d01d5_spline}
\end{figure}

In place of the computationally expensive
 low-$\ell$ polarization pixel likelihood, we seek a fast but accurate approximation.
 The WMAP team has shown that most of the information in the power law $\Lambda$CDM
 parameter space lies in multipoles $\ell \sim 2-7$ as essentially an overall amplitude
 of power \cite{Larson:2010gs}.  
However, in the broader parameter
 space allowed by the PCs of $G'$, we find that a single amplitude
 is insufficient to describe the information content of the pixel likelihood.
 
Instead we 
  fit the likelihood function to 
 a two band approximation
 \begin{equation}
 p_{E i} = \left( {1\over  {\Delta \ell_i }}\sum_{\ell=\ell_{i\rm min}}^{\ell_{i \rm max}} {\ell (\ell+1) C_\ell^{EE} \over 2\pi } \right)^{1/2} \,,
 \end{equation}
 where the first band $i=1$ has $\ell_{1\rm min}=4$, $\ell_{1\rm max}=6$ and
 the second band $i=2$ has $\ell_{2\rm min}=\ell_{2\rm max}=8$ and $\Delta \ell_i
 = \ell_{i \rm max}-\ell_{i \rm min}+1$ normalizes the parameter to reflect the average bandpower.
 We find that the pixel likelihood is well approximated by a Gaussian in these two bands
 for the models under consideration
 \begin{equation}\label{eq:fit}
 -2 \ln {\cal L}_{\rm pol}^{\ell<24} \approx A  + 
( {\bf p}_E - \bar{\bf p}_E )^T {\bf C}^{-1} ( {\bf p}_E - \bar{\bf p}_E )
\end{equation}
with the parameters $p_{E1}=0.2614$ $\mu$K, $p_{E2}=0.01955$ $\mu$K, $A=1645.84$
\begin{equation}
{\bf C}^{-1} = \left(
\begin{array}{cc}
498.31 & -214.23 \\
-214.23 & 190.23 \\
\end{array}\right)  \mu {\rm K}^{-2} \,.
\end{equation}
In Fig.~\ref{plot:contour_2bands_pE456_pE8} we show the accuracy of the fit
compared with the pixel likelihood for both power law models and models
with additional 5 PCs of $G'$.    Note that the power law models
lie on a 1D curve in this space and can be well parameterized by a single amplitude
whereas the 5PC models do not.  In fact, in the 2 band space models with
low $p_{E1}$ and high $p_{E2}$ that populate a direction nearly orthogonal to the
power law models are more strongly constrained than the total power at low $\ell$ would
suggest.  
Since this approximation has trivial computational cost, the net improvement in speed
is approximately $\sim 5 N_{\rm core}$.

For the cases of interest, the approximation works remarkably well. As an example, we have run an MCMC chain
with the exact pixel likelihood for the 5PC chain with WMAP data only.   In Fig. \ref{plot:tau_m4_fullL_1d01d5_spline}, we compare 
the posterior probability distribution of the optical depth $\tau$ and $m_4$  using the full likelihood  and the approximation.
The difference between the pixel likelihood and the approximation for the $5$ PCs  maximum likelihood model with WMAP data is likewise negligible: $|2\Delta \ln {\cal L}|=0.05$.

We have also checked that the likelihood approximation remains valid to 10\% or better
in the $(2\Delta \ln {\cal L})^{1/2}$ significance of differences between models with
varying reionization history as parameterized by ionization principal components
\cite{Mortonson:2007hq}.

Larger differences can occur for models with sharp, order unity, initial power spectrum features
at the horizon scale.   These project onto the temperature and polarization spectra
differently and leads to qualitatively different results for the temperature-polarization
cross spectrum.   In this case one can get discrepancies of  order unity in 
$2\Delta \ln {\cal L}$ that err on the side of allowing discrepant models.
Even these differences can typically be taken into account via importance sampling at a much smaller computational cost than evaluating the exact pixel likelihood during the MCMC run itself.
\vfill

\bibliographystyle{arxiv_physrev}
\bibliography{PC}

\def\eprinttmppp@#1arXiv:@{#1}
\providecommand{\arxivlink[1]}{\href{http://arxiv.org/abs/#1}{arXiv:#1}}
\providecommand{\arxivlinknopre[1]}{\href{http://arxiv.org/abs/#1}{#1}}
\providecommand{\eprintmod}[1][XXXX.XXXX]{\IfSubStr{#1}{arXiv}{\arxivlinknopre%
{#1}}{\arxivlink{#1}}}
\providecommand{\adsurl}[1]{\href{#1}{ADS}}
\begin{thebibliography}{29}
\expandafter\ifx\csname natexlab\endcsname\relax\def\natexlab#1{#1}\fi
\expandafter\ifx\csname bibnamefont\endcsname\relax
  \def\bibnamefont#1{#1}\fi
\expandafter\ifx\csname bibfnamefont\endcsname\relax
  \def\bibfnamefont#1{#1}\fi
\expandafter\ifx\csname citenamefont\endcsname\relax
  \def\citenamefont#1{#1}\fi
\expandafter\ifx\csname url\endcsname\relax
  \def\url#1{\texttt{#1}}\fi
\expandafter\ifx\csname urlprefix\endcsname\relax\def\urlprefix{URL }\fi

\bibitem{Bennett:2003bz}
WMAP, C.~L. Bennett {\em et~al.},
\newblock Astrophys. J. Suppl. {\bf 148}, 1 (2003),
  [\eprintmod[astro-ph/0302207]].

\bibitem{Peiris:2003ff}
WMAP, H.~V. Peiris {\em et~al.},
\newblock Astrophys. J. Suppl. {\bf 148}, 213 (2003),
  [\eprintmod[astro-ph/0302225]].

\bibitem{Covi:2006ci}
L.~Covi, J.~Hamann, A.~Melchiorri, A.~Slosar and I.~Sorbera,
\newblock Phys. Rev. {\bf D74}, 083509 (2006), [\eprintmod[astro-ph/0606452]].

\bibitem{Hamann:2007pa}
J.~Hamann, L.~Covi, A.~Melchiorri and A.~Slosar,
\newblock Phys. Rev. {\bf D76}, 023503 (2007), [\eprintmod[astro-ph/0701380]].

\bibitem{Mortonson:2009qv}
M.~J. Mortonson, C.~Dvorkin, H.~V. Peiris and W.~Hu,
\newblock Phys. Rev. {\bf D79}, 103519 (2009), [\eprintmod[0903.4920]].

\bibitem{Pahud:2008ae}
C.~Pahud, M.~Kamionkowski and A.~R. Liddle,
\newblock Phys. Rev. {\bf D79}, 083503 (2009), [\eprintmod[0807.0322]].

\bibitem{Joy:2008qd}
M.~Joy, A.~Shafieloo, V.~Sahni and A.~A. Starobinsky,
\newblock JCAP {\bf 0906}, 028 (2009), [\eprintmod[0807.3334]].

\bibitem{Dvorkin:2009ne}
C.~Dvorkin and W.~Hu,
\newblock Phys. Rev. {\bf D81}, 023518 (2010), [\eprintmod[0910.2237]].

\bibitem{Hazra:2010ve}
D.~K. Hazra, M.~Aich, R.~K. Jain, L.~Sriramkumar and T.~Souradeep,
\newblock \eprintmod[1005.2175].

\bibitem{Stewart:2001cd}
E.~D. Stewart,
\newblock Phys. Rev. {\bf D65}, 103508 (2002), [\eprintmod[astro-ph/0110322]].

\bibitem{Choe:2004zg}
J.~Choe, J.-O. Gong and E.~D. Stewart,
\newblock JCAP {\bf 0407}, 012 (2004), [\eprintmod[hep-ph/0405155]].

\bibitem{Dodelson:2001sh}
S.~Dodelson and E.~Stewart,
\newblock Phys. Rev. {\bf D65}, 101301 (2002), [\eprintmod[astro-ph/0109354]].

\bibitem{Hu:2003vp}
W.~Hu and T.~Okamoto,
\newblock Phys. Rev. {\bf D69}, 043004 (2004), [\eprintmod[astro-ph/0308049]].

\bibitem{Leach:2005av}
S.~M. Leach,
\newblock Mon. Not. Roy. Astron. Soc. {\bf 372}, 646 (2006),
  [\eprintmod[astro-ph/0506390]].

\bibitem{Sealfon:2005em}
C.~Sealfon, L.~Verde and R.~Jimenez,
\newblock Phys. Rev. {\bf D72}, 103520 (2005), [\eprintmod[astro-ph/0506707]].

\bibitem{Paykari:2009ac}
P.~Paykari and A.~H. Jaffe,
\newblock Astrophys. J. {\bf 711}, 1 (2010), [\eprintmod[0902.4399]].

\bibitem{Nicholson:2009zj}
G.~Nicholson, C.~R. Contaldi and P.~Paykari,
\newblock JCAP {\bf 1001}, 016 (2010), [\eprintmod[0909.5092]].

\bibitem{Bridle:2003sa}
S.~L. Bridle, A.~M. Lewis, J.~Weller and G.~Efstathiou,
\newblock Mon. Not. Roy. Astron. Soc. {\bf 342}, L72 (2003),
  [\eprintmod[astro-ph/0302306]].

\bibitem{Peiris:2009wp}
H.~V. Peiris and L.~Verde,
\newblock Phys. Rev. {\bf D81}, 021302 (2010), [\eprintmod[0912.0268]].

\bibitem{Kadota:2005hv}
K.~Kadota, S.~Dodelson, W.~Hu and E.~D. Stewart,
\newblock Phys. Rev. {\bf D72}, 023510 (2005), [\eprintmod[astro-ph/0505158]].

\bibitem{Christensen:2001gj}
N.~Christensen, R.~Meyer, L.~Knox and B.~Luey,
\newblock Class. Quant. Grav. {\bf 18}, 2677 (2001),
  [\eprintmod[astro-ph/0103134]].

\bibitem{Kosowsky:2002zt}
A.~Kosowsky, M.~Milosavljevic and R.~Jimenez,
\newblock Phys. Rev. {\bf D66}, 063007 (2002), [\eprintmod[astro-ph/0206014]].

\bibitem{gelman/rubin}
A.~Gelman and D.~Rubin,
\newblock Statistical Science {\bf 7}, 452 (1992).

\bibitem{Lewis:2002ah}
A.~Lewis and S.~Bridle,
\newblock Phys. Rev. {\bf D66}, 103511 (2002), [\eprintmod[astro-ph/0205436]].

\bibitem{Larson:2010gs}
D.~Larson {\em et~al.},
\newblock \eprintmod[1001.4635].

\bibitem{Chiang:2009xsa}
H.~C. Chiang {\em et~al.},
\newblock Astrophys. J. {\bf 711}, 1123 (2010), [\eprintmod[0906.1181]].

\bibitem{SHOES}
A.~G. Riess {\em et~al.},
\newblock Astrophys. J. {\bf 699}, 539 (2009), [\eprintmod[0905.0695]].

\bibitem{Mortonson:2007hq}
M.~J. Mortonson and W.~Hu,
\newblock Astrophys. J. {\bf 672}, 737 (2008), [\eprintmod[0705.1132]],
\newblock see also \url{http://background.uchicago.edu/camb_rpc}.

\end{thebibliography}

\end{document}